\documentclass[proof]{WileyASNA-v1}
\usepackage{bm}
\usepackage{graphicx}
\usepackage{caption}
\usepackage{amsmath}
\usepackage{dblfloatfix}
\usepackage{multirow}
\usepackage{epstopdf}
\articletype{Article Type}

\begin{document}
\title{The Low-Frequency Spectra of Radio Pulsars}
\author[1,2]{Ting Yu*}
\author[1,3]{Zhong-li Zhang}
\author[4]{Hong-yu Gong}
\author[5]{Zhi-gang Wen}
\address[1]{\orgdiv{Shanghai Astronomical Observatory}, \orgname{Chinese Academy of Sciences}, \orgaddress{\state{Shanghai, 200030}, \country{China}}}
\address[2]{\orgdiv{University of Chinese Academy of Sciences}, \orgname{Chinese Academy of Sciences}, \orgaddress{\state{No.19, Yuquan Road, Beijing, 100049}, \country{China}}}
\address[3]{\orgdiv{State Key Laboratory of Radio Astronomy and Technology}, \orgname{Chinese Academy of Sciences}, \orgaddress{\state{Beijing, 100101}, \country{China}}}
\address[4]{\orgdiv{Institute of Astronomy and Information}, \orgname{Dali University}, \orgaddress{\state{Yunnan, Dali, 671003}, \country{China}}}
\address[5]{\orgdiv{Xinjiang Astronomical Observatory}, \orgname{Chinese Academy of Sciences}, \orgaddress{\state{150 Science 1-Street, Urumqi, Xinjiang, 830011}, \country{China}}}
\corres{*Ting Yu, Shanghai Astronomical Observatory, Chinese Academy of Sciences, Shanghai 200030, China. \\  \email{yuting@shao.ac.cn} \\
}

\abstract{
Low-frequency spectral studies of radio pulsars represent a key method for uncovering their emission mechanisms, magnetospheric structure, and signal interactions with the surrounding interstellar medium (ISM). In recent years, more next-generation low-frequency radio telescopes (e.g., LOFAR, LWA and MWA) have enriched the observational window below 350 MHz, enabling more detailed explorations of the ISM effects, such as absorption and scattering, resulting in diverse spectral behaviors observed across different pulsars. This paper reviews the morphology of pulsar radio spectra, advances in spectral modeling, and the key physical processes governing the low-frequency emission. Looking ahead, next-generation instruments such as SKA-Low - with their unprecedented sensitivity - are expected to resolve outstanding questions in pulsar emission processes, offering insights into the extreme physical regimes governing these exotic objects.} 

\maketitle

\section{Introduction}\label{sec1}

Pulsars are highly magnetized, rapidly rotating neutron stars that form from the remnants of massive stars that have undergone supernova explosions. Since the discovery of pulsars in 1967 \citep{Hewish1968}, they have remained central to astrophysical research in extreme physical conditions. Pulsars typically have masses ranging from $1.1$ to $2.2~\mathrm{M_{\odot}}$ \citep{Stairs2004,Ozel2016}, radii of approximately $10$ to $12$ km \citep{Lattimer2004}, and average densities that are 2 to 3 times the nuclear saturation density \citep{Lattimer1978}. Their surface magnetic fields span 7 to 15 orders of magnitude, from $10^{7}$ to $10^{15}$ G \citep{Thompson1996,Gao2013,Gao2015,Gao2021}, and their rotation periods vary from $1.4$ ms to $76$ s \citep{Hessels2006,Caleb2022}. These characteristics make pulsars natural laboratories for testing theories related to strong gravitational fields, ultra-strong magnetic fields, and the equations of state for dense matter.

However, debates regarding the emission mechanisms of pulsars persist. A key point of contention is that theoretical models such as synchrotron radiation and curvature radiation \citep{Gold1968} do not provide a unified explanation for the diverse emission properties and the complex evolution of pulse profiles. The radio spectrum serves as a crucial probe of these emission mechanisms, revealing particle acceleration processes, magnetospheric geometry, and the altitude of the emission zone through the distribution of flux density $S_{\nu}$ across different frequencies. However, two long-standing challenges hinder research due to effects from the interstellar medium (ISM) propagation, including dispersion, scattering, and free-free absorption (FFA) \citep{Rickett1969,Kassim1994}, as well as limitations in observational technology. First, most pulsars lack broadband, high-precision spectral data, which leads to significant measurement errors in spectral indices. For example, interstellar scintillation can cause flux density variations ranging from two to ten times \citep{Bell2016}. Second, the physical origins of the diversity in spectral shapes (such as simple power-law, broken power-law, log-parabolic spectrum, low-frequency turnover, high-frequency cutoff \citep{Jankowski2018}) remain unresolved, highlighting the imperative need to integrate emission mechanisms and environmental effects into a coherent explanatory framework.

Early studies established the foundational paradigms for spectral analysis. In 1968, \citet{Robinson1968} revealed non-thermal power-law spectral characteristics from multi-frequency observations of CP 1919. Subsequently, \citet{Sieber1973} discovered low- and high-frequency turnover features and introduced the concept of a critical frequency $\nu_{\rm c}$. However, progress in spectral classification has been constrained by limited frequency coverage and low sensitivity, resulting in much of the analysis remaining at a statistical level. In a subsequent statistical study of the low-frequency flux densities of 52 pulsars, \citet{Izvekova1981} found that 44 pulsars show low-frequency spectral turnover, where the flux density peaks at a characteristic turnover frequency of $\nu_{\rm m} = 130 \pm 80$ MHz. 

Since the 21st century, advancements in low-frequency radio technologies, such as UTR-2, LWA, LOFAR, and MWA, have enriched observational data in the 10$\sim$350 MHz range, enabling more detailed studies of pulsars and the ISM \citep{Zakharenko2013,Stovall2015,Bilous2016,Bhat2023}. The broad frequency coverage has also revealed correlations between spectral shapes and factors such as spin-down energy loss rate, age, and the interstellar environment. For instance, millisecond pulsars (MSPs) exhibit a significantly steeper average spectral index of $-1.9 \pm 0.1$, compared to $-1.72 \pm 0.04$ for normal pulsars \citep{Toscano1998}. Furthermore, the turnover frequency $\nu_{\rm m}$ for pulsars located within supernova remnants (SNRs) can reach up to 1 GHz, significantly exceeding the 300 MHz typical of isolated pulsars \citep{Kijak2011}.

The observed spectral indices of pulsars range from $-4$ to $0$. The average spectral index for the sample of \citet{Lorimer1995} is $-1.6$, while \citet{Maron2000}, based on \citet{Lorimer1995}'s research, reports an average spectral index of $-1.8 \pm 0.2$ for frequencies above 100 MHz. Considering the selection bias and incomplete data, a Gaussian distribution for the spectral indices was modeled. \citet{Bates2013} simulations suggest a potential spectral index of $-1.4$. The \texttt{pulsar\_spectra}\footnote{https://github.com/NickSwainston/pulsar\_spectra} software library, developed by \citet{Swainston2022}, analyzed the spectral characteristics of 886 pulsars, finding an average spectral index of $-1.64 \pm 0.80$ for power-law modeled pulsars, consistent with previous studies \citep{Xu2024}.

The necessity for current research arises from three main scientific demands: First, steep spectral characteristics result in higher low-frequency flux densities, which facilitates the detection of faint pulsars. Second, next-generation facilities (such as Square Kilometre Array-Low, SKA-Low) are set to achieve nanojansky sensitivity with continuous coverage from hundreds of MHz to several GHz \citep{Keane2015}, necessitating a systematic spectral classification framework to provide benchmarks for large-scale data mining. Finally, pulsar timing arrays depend on precise ISM corrections to enhance gravitational wave detection accuracy, and the evolution of spectra contains crucial information regarding electron density fluctuations \citep{Bilous2020}. However, existing spectral databases still suffer from significant sample biases: low-frequency surveys are heavily concentrated in the Northern hemisphere, high-energy pulsars (such as gamma-ray pulsars) often have insufficient radio spectral coverage, and most pulsars lack multi-epoch observations to differentiate their inherent spectral characteristics from transient interstellar scintillation effects. This paper aims to systematically review the progress in radio pulsar spectral research, analyze the interplay between morphological classification, emission mechanisms, and medium effects, and explore how multi-messenger observations in the SKA era can address the long-standing challenges in pulsar radiation.

\section{Classification of Spectral Morphologies}\label{sec2}

The diversity in the radio spectra of pulsars directly reflects the interaction between emission mechanisms and the interstellar environment. This section systematically reviews the observational patterns and physical explanations of low- and high-frequency turnover spectra, as well as high-frequency cutoff spectra, based on frequency-dependent characteristics.

\subsection{Low-Frequency Spectral Turnover}

Low-frequency spectral features in pulsars are attributed to a combination of intrinsic emission processes and propagation effects. In an early study, \citet{Slee1986} analyzed pulsar spectra in the 80$\sim$1400 MHz band and found predominantly steep spectral indices, although no significant correlation emerged between these spectral indices and other pulsar parameters. Subsequent investigations into the low-frequency emission of radio pulsars highlighted the influence of the ISM. Notably, \citet{Rankin1970} was the first to apply an ISM model when analyzing observations of the Crab pulsar (PSR B0531+21), revealing that its spectral index is approximately $-2.9 \pm 0.4$ above 150~MHz and shows a sharp turnover at a critical frequency of $\nu_{\rm m} \approx 100$~MHz. Moreover, the observed pulse arrival times were found to be delayed in a manner consistent with dispersion in a tenuous plasma.

Subsequently, \citet{Malofeev1994} employed wide-band observations of 45 pulsars and classified them into two categories: those that follow a simple power-law and those that undergo spectral turnover at low frequencies, consistent with predictions from coherently excited curvature radiation models.

Further observations suggest that MSPs generally do not exhibit pronounced low-frequency turnover in the 102$\sim$408 MHz range \citep{Kramer1998,Malofeev2000}, in contrast to normal pulsars. This discrepancy may arise from the more complex magnetospheric structures (e.g., non-dipolar field configurations) or relativistic particle flows in MSPs, which could lead to a radially compressed emission region and suppress the notable FFA commonly invoked at low radio frequencies \citep{Kuzmin2001}. Nonetheless, \citet{Kuniyoshi2015} performed a statistical analysis with VLSSr, WENSS, and NVSS catalogs and found that some MSPs might indeed show low-frequency turnover. Furthermore, MSP observations indicate that changes in pulse profile amplitude ratios often dominate the low-frequency regime, likely reflecting the influence of aberration and retardation effects. This in turn suggests a highly compressed magnetospheric emission region in MSPs \citep{Kondratiev2016}.

\begin{figure}[!htbp]
    \centering
    \includegraphics[width=8cm]{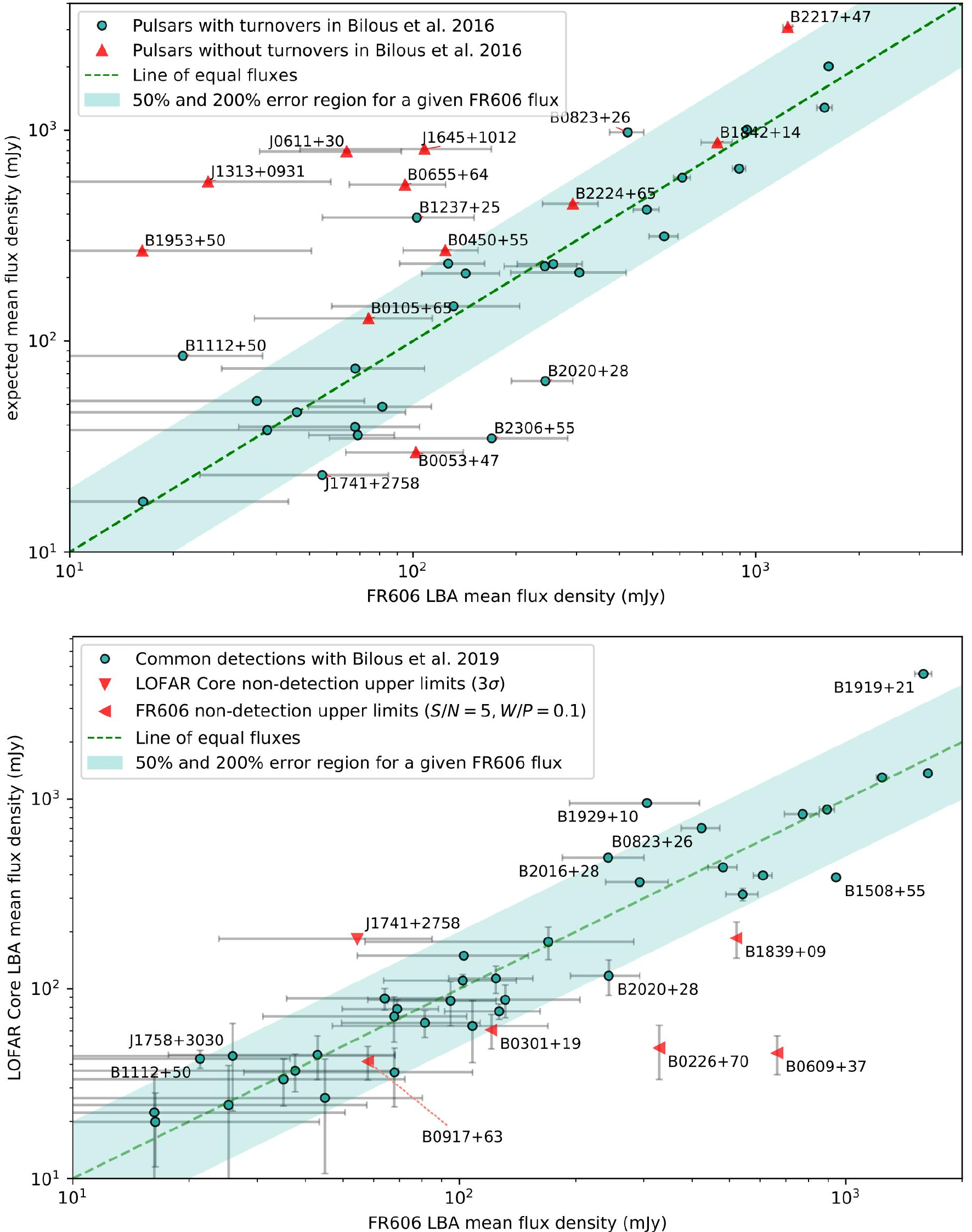}
    \caption{Top: Flux densities from \citet{Bondonneau2020} (25$\sim$80 MHz) compared to extrapolated values from high-frequency spectral indices \citep{Bilous2016}. Blue dots indicate pulsars with known spectral turnovers, while red triangles represent those modeled with a simple power law.\\
    Bottom: Consistency check between \citet{Bondonneau2020} and LOFAR Core LBA \citep{Bilous2020} measurements, with green lines and shaded regions indicating systematic uncertainties. Red triangles show upper limits for pulsars detected by only one instrument.}
    \label{LOFAR2020}
\end{figure}

The idea that low-frequency turnover is common among pulsars has been supported by an expanded sample of flux density measurements. For instance, \citet{Stovall2015} observed additional pulsars and found that many exhibit spectral turnover at low frequencies.  \citet{Bondonneau2020}, utilizing LOFAR observations at the FR606 French station below 100 MHz, confirmed such turnover in five pulsars and hypothesized that it may be a broader characteristic of the pulsar population as a whole (see Fig. \ref{LOFAR2020}).

The emergence of this phenomenon is not coincidental; the low-frequency spectral characteristics of pulsars may be subject to various propagation effects. \citet{Pilia2016} reported pulse broadening at 40 MHz, consistent with radius-to-frequency mapping (RFM) and birefringence theories, reinforcing the importance of propagation effects in shaping the low-frequency spectral profile. Overall, these results highlight the role of geometric and magnetic field complexities \citep{Li2023,Li2024}, as well as variations in the local ISM environment, in driving the observed low-frequency turnover.

\subsection{High-Frequency Spectral Turnover}

High-frequency emission properties are a key topic in pulsar physics. The complex spectral evolution in the high-frequency regime ($>$1~GHz) reflects both the diversity of underlying radiation mechanisms and the interactions with the pulsar's immediate environment and the ISM. Observational and theoretical studies have identified pulsars that exhibit high-frequency turnover, commonly known as "Gigahertz-Peaked Spectrum (GPS)" pulsars \citep{Kijak2009}.

Systematic investigations into high-frequency turnover behavior were initiated by the seminal works of \citet{Kijak2004}, who reported high-frequency turnovers in two pulsars, with turnover frequencies exceeding 1~GHz and around 600~MHz, respectively. This suggests that a pulsar's local interstellar environment can modulate the turnover frequency \citep{Kijak2007}. Later, \citet{Kijak2009,Kijak2011} noted that pulsars within SNRs or other extreme environments may experience FFA in the surrounding material or interstellar medium, producing spectral peaks near 1~GHz. Based on broadband observations, \citet{Tarczewski2012} identified 22 candidate GPS pulsars. 

Analyzing PSR B1800-21, \citet{Basu2016} confirmed that FFA satisfactorily explains its spectral turnover. Building on this, \citet{Rajwade2016} used the GMRT to image five newly discovered GPS pulsars and, by incorporating FFA models, were the first to constrain the electron density, temperature, and geometric scale of the absorbing medium. They suggested that these absorption regions might be associated with SNRs or pulsar wind nebulae. Furthermore, \citet{Basu2018} conducted extensive multi-frequency observations of six pulsars embedded in pulsar wind nebulae, four of which displayed GPS features. The inferred absorption parameters were consistent with the spatial distribution of ionized gas in the nebula. These efforts not only highlight the dominant role of FFA in high-frequency turnover phenomena but also provide critical insights for future quantitative investigations of environmental properties.

\begin{figure}[!htbp]
    \centering
    \includegraphics[width=8cm]{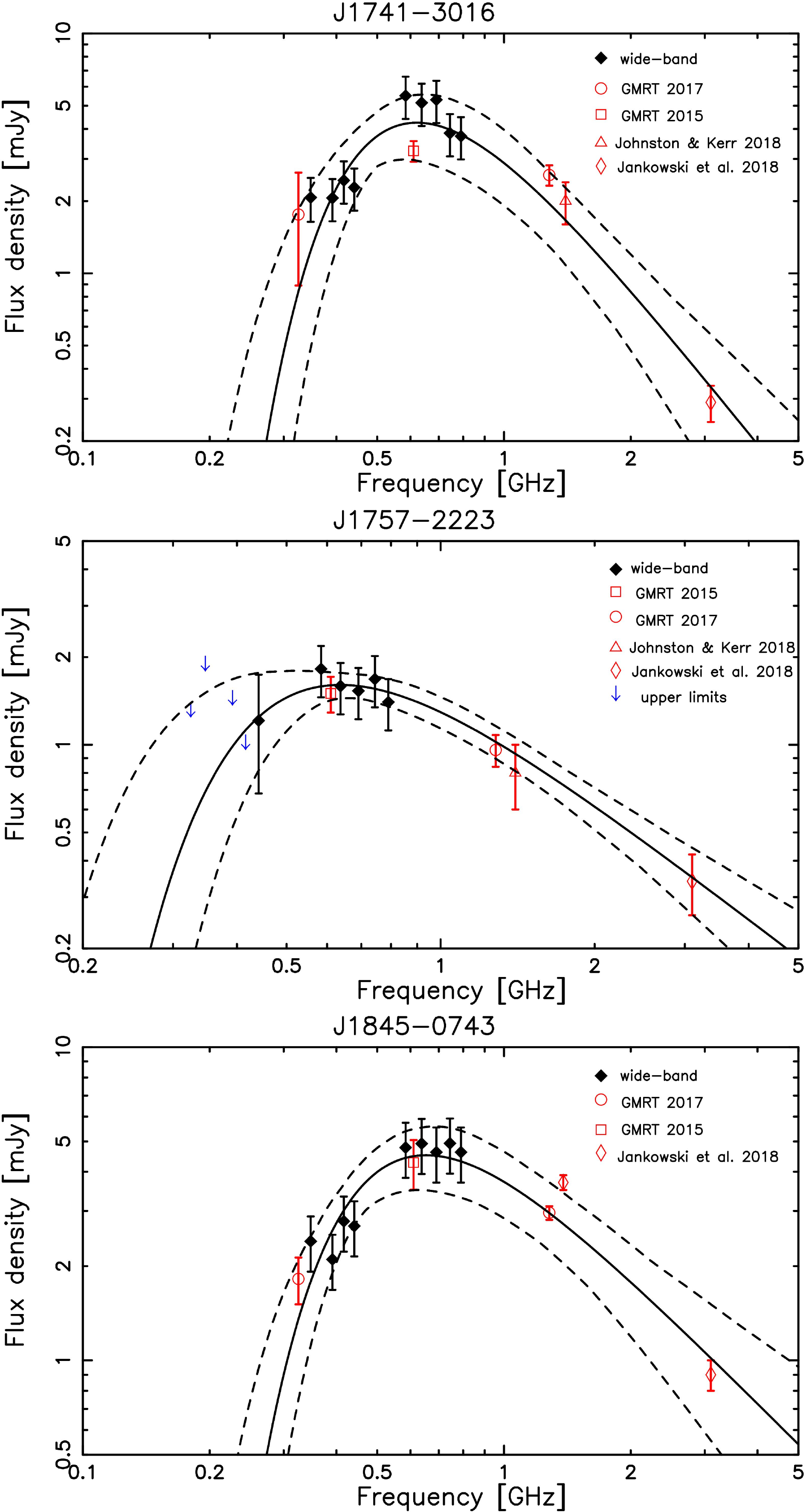}
    \caption{This figure illustrates the typical characteristics of GPS pulsars, with their spectral nature approximated using a free-free thermal absorption model (solid dark line) and the corresponding 1-$\sigma$ envelopes for the model fits (dotted lines) \cite{Rozko2021}.}
    \label{GPS}
\end{figure}

\citet{Kijak2021} reported six additional turnover pulsars at 325~MHz using the GMRT, increasing the total number of known GPS pulsars to 33. Their results showed that GPS pulsars are often spatially associated with SNRs or H II regions, and that absorption strength correlates positively with local electron density gradients. \citet{Rozko2021} using the GMRT have identified three new gigahertz-peaked spectra pulsars, revealing frequency turnovers at 620 MHz, 640 MHz, and 650 MHz, enhancing the accuracy of spectral modeling (see Fig. \ref{GPS}).

Recent work by \citet{Xu2024} confirmed that 67\% of the 22 GPS pulsars exhibit high dispersion measures (DM $> 150$ pc cm$^{-3}$), providing direct validation for the "high DM selection effect" hypothesis—suggesting that due to the complex inhomogeneity of the ISM, the DM distribution can indirectly reflect the environmental characteristics of different types of absorption pulsars.

\subsection{High-Frequency Cutoff Spectrum}

Early high-frequency observations (1.4$\sim$10.7~GHz) of pulsar emission revealed deviations from a simple power-law decay. In a large-scale study of 183 pulsar pulse profiles, \citet{Seiradakis1995} found significant departures from the predictions of curvature and synchrotron models, suggesting that the underlying physical processes at these frequencies are not well understood. Later, \citet{Kijak1998} conducted broadband observations (1.4$\sim$5~GHz) and determined an average spectral index of $\alpha \approx -1.9$, indicating a steep decline in high-frequency emission efficiency with increasing frequency. However, the cause of this steep slope remains unclear and requires further investigation.

From a geometrical perspective, \citet{Ommen1997} reconstructed the three-dimensional structure of pulsar emission beams using polarization data, suggesting that geometric effects along the beam’s path may cause high-frequency emission cutoffs. Meanwhile, \citet{Kramer1999} observed that MSPs exhibit stronger emission extending into the GHz range. This behavior is attributed to higher Lorentz factors ($\gamma$) or smaller curvature radii ($\rho$) in their magnetospheres, resulting in distinct high-frequency evolution compared to normal pulsars.

\begin{figure}[!htbp]
    \centering
    \includegraphics[width=8cm]{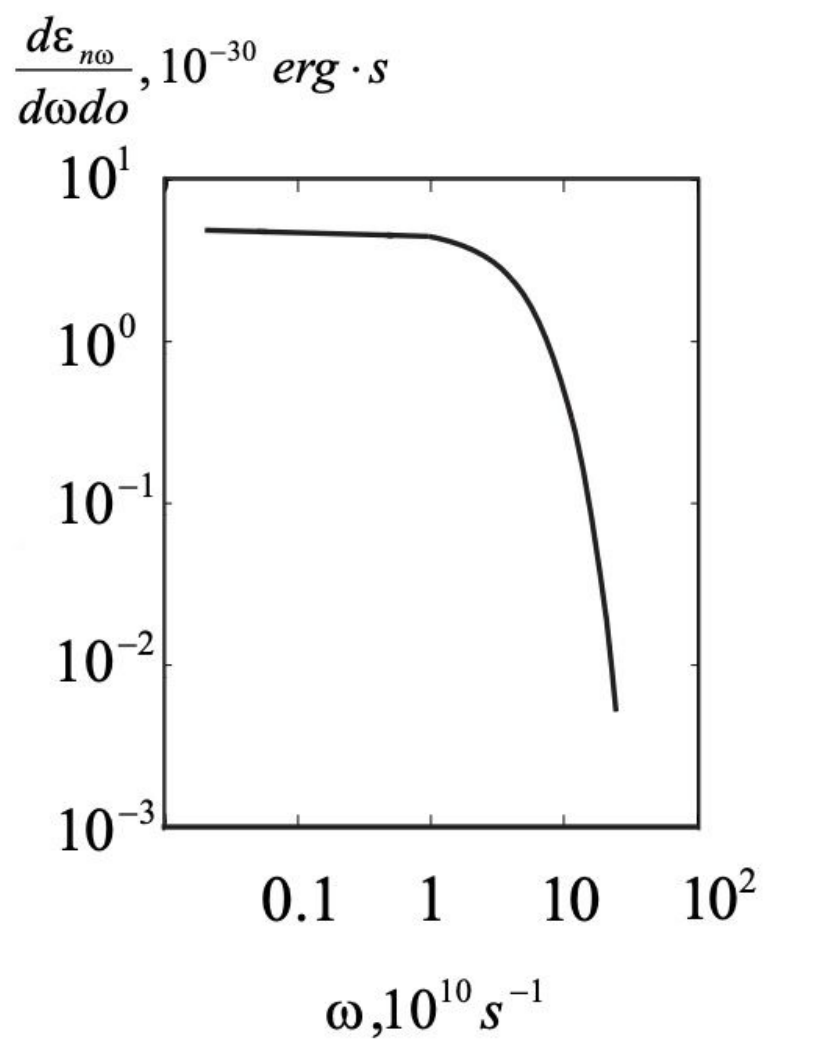}
    \caption{The emission spectrum of a particle undergoing acceleration in an electric field that increases linearly with altitude above the surface of the pulsar \citep{Kontorovich2013}.}
    \label{high_cutoff}
\end{figure}

\citet{Kontorovich2013} propose a new theoretical model based on electron acceleration in the polar gap to explain the high-frequency cutoff in pulsar radio emission. The model suggests that as electrons are accelerated in the polar gap, their acceleration increases initially but decreases as their velocity approaches the relativistic limit. The cutoff frequency is given by $\omega _{cf} \approx \pi \sqrt {2eE_{0} /mh} $
(Fig. \ref{high_cutoff}), where $E_{0}$ is the electric field and $h$ is the height of the gap.

Another theoretical explanation for the high-frequency cutoff is the coherent curvature radiation from dynamically emitting electron bunches in the magnetosphere. The $\omega _{cf}$ is determined by the interplay between the classical peak frequency of curvature radiation ($\omega_{\text{peak}}$, $\frac{2\gamma^2}{\tau_B}$) and the bunch lifetime ($\tau_B$). However, the sensitivity of current radio facilities at submillimeter to terahertz frequencies is limited, making direct searches for high-frequency cutoffs challenging. Progress in this area will rely on multiwavelength observational campaigns and refined magnetospheric simulations to gain deeper insights into dynamically fluctuating charged bunches in magnetospheres \citep{Yang2023}.

\section{Spectral Models}\label{spectra}

Modeling pulsar radio spectra involves both the intrinsic emission mechanisms and the extrinsic effects of interstellar propagation. This section categorizes existing models into empirical fitting and theoretical physical models, with a focus on competing explanations for low-frequency turnover phenomena.

\subsection{Empirical Formulations}

Empirical formulations use general mathematical functions (as shown in Fig. \ref{spectra}) to characterize spectral shapes, without relying on assumptions about specific physical mechanisms. Such models exhibit considerable flexibility and robustness in fitting actual observational spectra \citep{Jankowski2018,Swainston2022}.
 
\begin{figure}[!htbp]
    \centering
    \includegraphics[width=9cm]{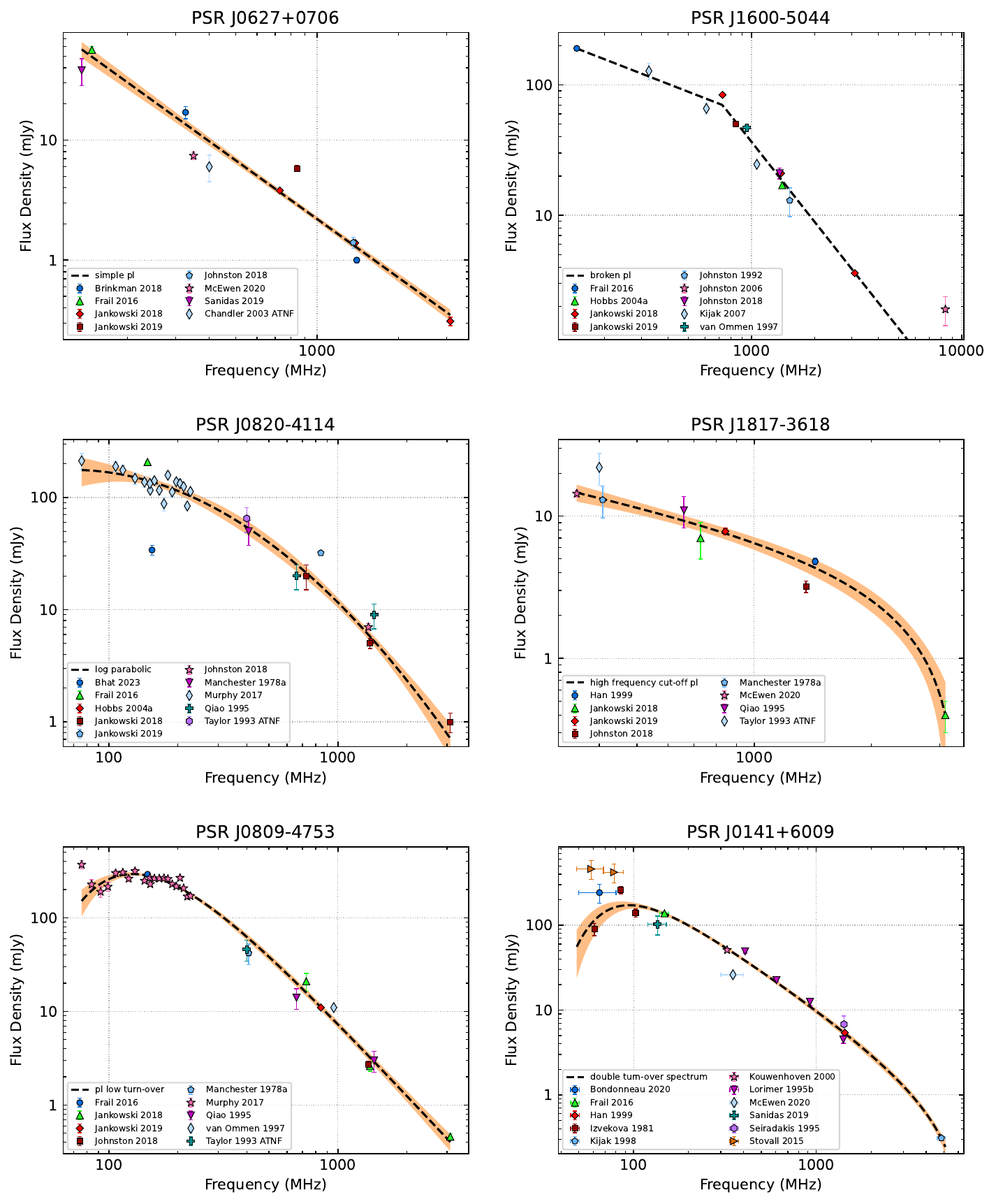}
    \caption{Spectral fitting of pulsars showing different morphological types.}
    \label{spectra}
\end{figure}

\subsubsection{Simple Power-Law Spectrum}\label{spl}

Pulsars dominated by non-thermal radiation typically exhibit a power-law spectrum, where the flux density is given by:
\begin{equation}
S_\nu = b\left(\frac{\nu}{\nu_0}\right)^{\alpha}.
\end{equation}
Here, $b$ is the intrinsic flux scaling factor, $\nu_0$ is the reference frequency, and $\alpha$ is the spectral index.

\subsubsection{Broken Power-Law Spectrum}\label{bpl}

When significant bends in the spectrum are observed, a broken power-law model is introduced:
\begin{equation}
S_\nu = b \begin{cases}
\left(\frac{\nu}{\nu_0}\right)^{\alpha_1}, & \nu \leq \nu_{\rm b} \\
\left(\frac{\nu}{\nu_0}\right)^{\alpha_2}\left(\frac{\nu_{\rm b}}{\nu_0}\right)^{\alpha_1-\alpha_2}, & \nu > \nu_{\rm b}
\end{cases}
\end{equation}
where $\nu_{\rm b}$ is the break frequency, and $\alpha_1$ and $\alpha_2$ are the spectral indices for the low and high-frequency bands, respectively. 

\subsubsection{Log-Parabolic Spectrum}\label{lps}

Curved spectra can be described using a log-parabolic function:
\begin{equation}
\log_{10}S_\nu = a\left[\log_{10}\left(\frac{\nu}{\nu_0}\right)\right]^2 + b\log_{10}\left(\frac{\nu}{\nu_0}\right) + c.
\end{equation}
In this expression, the curvature parameter $a$ characterizes the peak position of the spectrum. $b$ is the
spectral index for $a = 0$ and $c$ is a constant. When $a > 0$, it corresponds to GPS pulsars. This spectra may indicate the continuous acceleration of particle energy distributions.

\subsubsection{High-Frequency Cutoff Power-Law}\label{hfcpl}

Some pulsars show a sharp decline in radiation beyond a critical frequency $\nu_{\rm c}$, described by the model:
\begin{equation}
S_\nu = b\left(\frac{\nu}{\nu_0}\right)^{-2}\left(1 - \frac{\nu}{\nu_{\rm c}}\right), \quad (\nu < \nu_{\rm c}).
\end{equation}
The cutoff frequency $\nu_{\rm c}$ is related to the surface magnetic field strength $B$ and the period $P$ as follows:
\begin{equation}
\nu_{\rm c} = \sqrt{\frac{\pi e B}{m_{\rm e} c P}},
\end{equation}
where $e$ denotes the elementary charge, $m_{\rm e}$ is the mass of the electron, and $c$ is the speed of light in vacuum \citep{Kontorovich2013}.

\subsubsection{Low-Frequency Turnover Spectrum}\label{lftos}

The power-law model for low-frequency turnover is expressed as:

\begin{equation}
S_\nu = b \cdot \left(\frac{\nu}{\nu_0}\right)^\alpha \exp\left[ \frac{\alpha}{\beta} \left( \frac{\nu}{\nu_{\rm c}}\right)^{- \beta}\right], 
\end{equation}
where $\nu_{\rm c}$ is the turnover frequency and $\beta$ ($0 < \beta \leq 2.1$) controls the smoothness of the turnover. The fitting parameters for this model are $\alpha$, $\nu_{\rm c}$, $b$, and $\beta$. This model is based on the process of SSA (details in Section \ref{sec:ssa}), which is a primary mechanism for the low-frequency turnover phenomenon \citep{Sieber1973}. Furthermore, when $\beta = 2.1$, this model is equivalent to the thermal FFA model (details in Section \ref{sec:ffa}) \citep{Kijak2017}.

\subsubsection{Double turnover Spectrum}\label{dts}

Double turnover spectrum, characterized by both a low-frequency turnover and a high-frequency cutoff power law:

\begin{equation}
S_\nu = b \left( \frac{\nu}{\nu_0} \right)^a \left( 1 - \frac{\nu}{\nu_{\rm c}} \right) \exp \left[ \frac{a}{\beta} \left( \frac{\nu}{\nu_{\text{peak}}} \right)^{-\beta} \right], \quad \nu < \nu_{\rm c}.
\end{equation}

\subsection{Theoretical Turnover Physical Models}

The physical model links spectral features to astrophysical parameters by examining radiation mechanisms and propagation effects. Pulsar radiation mechanisms are tied to particle acceleration processes, with electron energy spectra commonly assumed to follow a power-law distribution:
\begin{equation}
N(E) = k E^{-p},
\end{equation}
where $p$ is the electron energy spectral index. Variations in the particle spectrum influence the synchrotron radiation photon distribution. For synchrotron radiation, the spectral index $\alpha$ relates to $p$ as $\alpha = \frac{p - 1}{2}$ \citep{Rybicki1979}, while for curvature radiation, the relationship is $\alpha = \frac{p - 2}{3}$ \citep{Cheng1986}. Fitting observed pulsar spectral indices $\alpha$ generally reveals a power-law distribution for synchrotron radiation (as detailed in Section \ref{spl}).

\subsubsection{Synchrotron Self-Absorption}\label{sec:ssa}

The synchrotron self-absorption (SSA) model assumes that the radiating particles follow a power-law energy spectrum distribution.
According to standard theoretical analysis \citep{Pacholczyk1970}, for particles that meet the power-law energy spectrum distribution, the emission coefficient $j_\nu$ and the absorption coefficient $\alpha_\nu$ can be represented as:
\begin{equation}
j_\nu = \frac{k B_\perp}{(p + 1)} \Gamma\left( \frac{3p + 19}{12} \right) \Gamma\left( \frac{3p - 1}{12} \right) \left(\frac{\nu}{B_\perp}\right)^{-\frac{(p-1)}{2}},
\end{equation}
\begin{equation}
\alpha_\nu = k B_\perp^{(p + 2)/2} \Gamma\left(\frac{3p + 2}{12}\right) \Gamma\left(\frac{3p + 22}{12}\right) \nu^{-(p + 4)/2}.
\end{equation}

Assuming that the source function of the radiation $S_\nu$ is constant, i.e., $S_\nu = \frac{j_\nu}{\alpha_\nu}$, we derive the standard synchrotron radiation spectrum:
\begin{equation}
    S_\nu \propto B_{\perp}^{-1/2} \nu^{5/2} \left\{ 1 - \exp\left[-\left(\frac{\nu}{\nu_1}\right)^{-(p + 4)/2}\right] \right\},
\end{equation}
where the SSA frequency $\nu_1$ is defined as the frequency at which the optical depth satisfies $\tau(\nu_1) = 1$, marking the critical frequency that separates the optically thick and optically thin regions of synchrotron radiation.

Under the assumption that the optical depth $\tau$ in the emission region is constant, the SSA frequency is given by:
\begin{equation}
    \nu_1 \propto (sk)^{2/(p+4)} B_\perp^{(p+2)/(p+4)},
\end{equation}
where $s$ is the scale of the radiation region, $B_\perp = B \sin \theta$, $B$ is the magnetic field strength, and $\theta$ is the pitch angle of the spiraling radiating particles \citep{O'Connor2005}.

According to \citet{Sieber1973}'s approximation for the SSA model for incoherent radiation and small magnetic fields, we have:
\begin{equation}
S_\nu = b_1 \left(\frac{\nu}{\nu_0}\right)^{2.5} \left(1 - e^{-b_2 \nu^{\alpha - 2.5}}\right),
\end{equation}
where $b_1$ is the scaling factor for the intrinsic flux of the pulsar, and $b_2$ is a constant coefficient.

\subsubsection{Coherent Curvature Radiation}\label{sec:ccr}

\citet{Ochelkov1984} proposed a model of coherent curvature radiation to explain the low-frequency turnover, where the nonuniformity and anisotropy of the radiation source give rise to two characteristic frequencies in the spectrum:

\begin{equation}
\nu_{\rm b} = \left[\frac{3}{4} \left(\frac{2}{3}\right)^{2/3} \frac{(k-2)e^{2}\theta_{\rm p}^{2(k-2)}\dot{N}}{m_{\rm e}cR^{2}\Gamma} \left(\frac{8}{9} \frac{R}{c\Gamma^{3}}\right)^{-k} \right]^{1/(2+k)},
\end{equation}
which is determined by the plasma parameters and the curvature radiation mechanism.

\begin{equation}
\nu_{\text{peak}} = \nu_{\rm b} \theta_{\rm p}^{\left[\frac{16}{3(2+k)}\right]}.
\end{equation}
This frequency is related to the polar cap angle $\theta_p$, and varies with the changes in the polar cap region as the pulsar rotates. Here, $k$ is a constant; $e$ and $m_e$ are the charge and mass of an electron, respectively; $\dot{N}$ is the number of particles emitted from the pulsar per unit time; $c$ is the speed of light; $\Gamma$ is the Lorentz factor; $R$ is the radius of the neutron star; and $\theta_{\rm p}$ is defined as $R\Omega/c$, where $\Omega = 2\pi/P$ is the angular velocity of the pulsar.

Prior to $\nu_{\text{peak}}$, the spectrum follows:
\begin{equation}
S_{\nu} \propto \nu^{-\frac{4(k-2)}{6+k}},
\end{equation}
After $\nu_{\text{peak}}$, the spectrum becomes steeper:
\begin{equation}
S_{\nu} \propto \nu^{-(k-2)}.
\end{equation}

It has been found that when $k$ is set to 4 and 5 for normal pulsars and MSPs, respectively, the model predictions align well with the occurrence of turnover and cutoff phenomena in both low and high-frequency regions \citep{Xu2024}.

\subsubsection{Free-Free Absorption}\label{sec:ffa}

The free-free absorption (FFA) model is the process in which photons collide with free electrons and ions in the ISM. When an electron absorbs the energy of a photon, it transitions from a low-energy free state to a high-energy free state, leading to the suppression of the radiative flux in specific frequency bands. This mechanism is one of the core physical processes explaining the turnover phenomenon in pulsar spectra.

\citet{Kijak2017} simulated the spectra of all known pulsars based on the hypothesis that the spectral turnover is caused by free-free thermal absorption in the ISM. They constrained the physical parameters of the absorbing medium through observations, thereby distinguishing possible pulsars of absorption. In this model, it is assumed that the intrinsic spectrum of the pulsar follows a simple power-law form. To estimate the optical thickness, an approximate formula for free-free thermal absorption is employed:

\begin{equation}
S(\nu) = b\left(\frac{\nu}{\nu_0}\right)^{\alpha} e^{-B\nu^{-2.1}},
\end{equation}
where $b$ is the scaling factor for the intrinsic flux of the pulsar, $\alpha$ is the intrinsic spectral index, and the frequency $\nu$ is expressed in GHz. The optical depth $\tau$ is given by the product of a frequency-dependent factor and a frequency-independent parameter, $B = 0.08235 \times T_{\rm e}^{-1.35} EM$, 
where $T_{\rm e}$ represents the electron temperature, and $EM$ denotes the emission measure (in $\mathrm{cm}^{-6}\,\mathrm{pc}$).

The free-free thermal absorption model proposed by \citet{Sieber1973} is described as:

\begin{equation}
S_\nu = b\left(\frac{\nu}{\nu_0}\right)^{\alpha} e^{-\tau}
\end{equation}

\begin{equation}
\tau = 3.014 \times 10^{4} \times EM \times T^{-1.5} \nu^{-2} \times \ln(49.55 \times T^{1.5} \nu^{-1}),
\end{equation}
where the unit of $\nu$ is MHz, $T$ is the temperature of the absorbing cloud (in K), and the definition of $EM$ is consistent with that in the \citet{Kijak2017} model.

\section{Statistical Analysis of Spectrum Turnover in \texttt{pulsar\_spectra}}

\begin{figure}[!htbp]
    \centering
    \includegraphics[width=8cm]{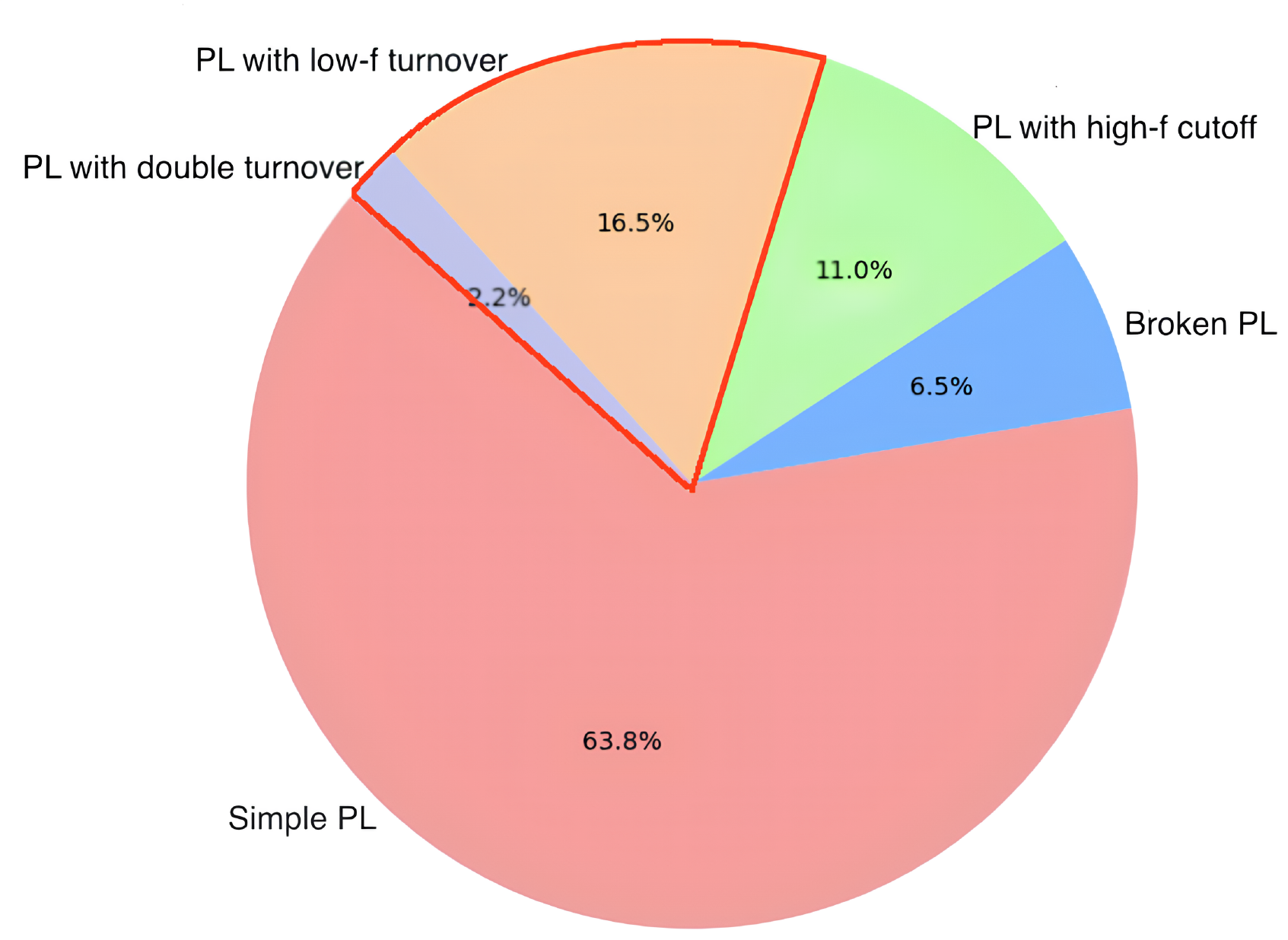}
    \caption{The distribution of the spectral composition in different forms based on \texttt{pulsar\_spectra}.}
    \label{distribution}
\end{figure}

This study presents a comprehensive analysis of pulsar radio spectra using the spectral fitting tool \texttt{pulsar\_spectra}, developed by \citet{Swainston2022}. The results indicate that approximately 64\% of the pulsars exhibit a simple power-law spectral characteristic, described by $S_\nu \propto \nu^\alpha$, while 11\% display high-frequency cutoff features. Additionally, about 20\% of the pulsars show low-frequency turnover spectra, which include both simple and double turnover forms, as illustrated in Fig. \ref{distribution}. It is important to emphasize that, due to the data being compiled from multiple epochs and observational instruments, systematic errors in flux density measurements—such as calibration biases and variations in instrument sensitivity—could significantly influence the statistical outcomes.

\begin{figure*}[!ht]
\begin{center}
    \includegraphics[scale=0.35, angle=0]{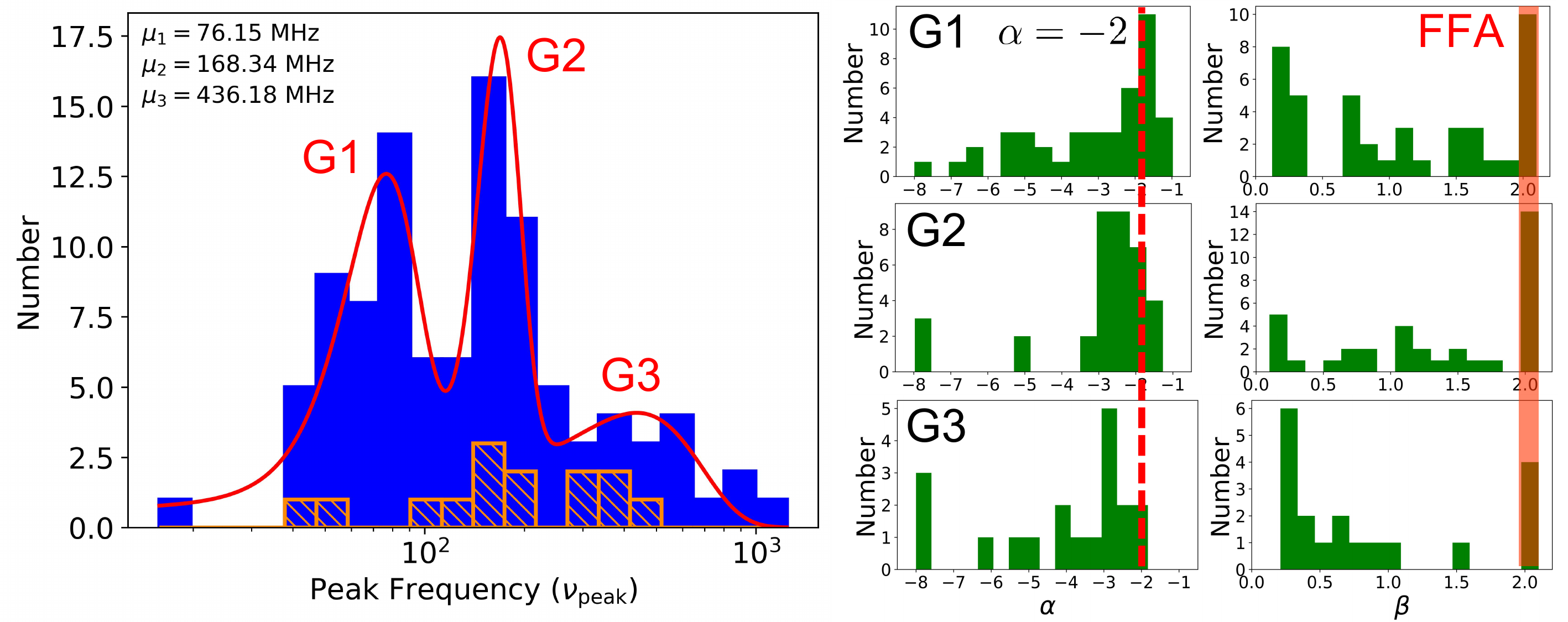}
    \caption{Statistical results of low-frequency turnover pulsars.}
    \label{statistics}
\end{center}
\end{figure*}

We conducted statistical modeling and spatial association analysis on 98 selected low-frequency turnover pulsars. As shown in the left panel of Fig. \ref{statistics}, the turnover peak frequency $\nu_{\text{peak}}$ exhibits a tri-modal distribution:

\begin{itemize}
    \item \textbf{Group G1} ($\nu_{\text{peak}} < 100 \text{ MHz}$): This group comprises 43.9\% of the turnover pulsars. The statistical significance of this group needs further verification due to potential biases in observations at extremely low frequencies ($\nu < 300 \text{ MHz}$) and interstellar scattering effects.
    
    \item \textbf{Group G2} ($\nu_{\text{peak}} \approx 170 \text{ MHz}$): This group accounts for 36.7\%, representing the most statistically significant sample.
    
    \item \textbf{Group G3} ($\nu_{\text{peak}} \approx 450 \text{ MHz}$): This group includes 19.4\% of the turnover pulsars, some of which are low-frequency thermal absorption GPS pulsars reported in the literature \cite{Kijak2021}.
\end{itemize}

The parameter distribution presented in the right panel of Fig. \ref{statistics} indicates a significant negative correlation between $\nu_{\text{peak}}$ and the power-law index $\alpha$; specifically, as $\nu_{\text{peak}}$ increases, the steepness of the spectrum also increases. When $\beta = 2.1$, the model reduces to the classical FFA spectrum, where the optical depth $\tau \propto \nu^{-2.1}$ aligns with the absorption theory of a thermal electron medium (see Section \ref{sec:ffa}). This parameter threshold serves as a crucial criterion for distinguishing between FFA and SSA mechanisms.

Further environmental surveys of Groups G2 and G3 revealed that 15 pulsars are spatially correlated with dense plasma environments (such as SNRs, pulsar wind nebulae, the Gum Nebula, Loop I, and Local Bubble interfaces). Notably, the proportion of FFA-dominated pulsars in Group G2 reached 39\%, significantly higher than that in Groups G1 (23\%) and G3 (21\%). Some high-frequency turnover pulsars in Group G3 ($\nu_{\text{peak}} > 1 \text{ GHz}$) may exhibit physical consistency with the previously reported GPS pulsars, while the low reliability of Group G1 can be attributed to several factors:

\begin{itemize}
    \item Calibration uncertainties in flux density measurements at extremely low frequencies ($\nu < 100 \text{ MHz}$);
    \item Spectral distortions caused by strong scattering broadening;
    \item Temporal effects due to the ISM refraction and scintillation.
\end{itemize}

\section{Summary and Discussion}

This paper provides a comprehensive review of the advancements in the study of pulsar radio pulse spectra, with a particular emphasis on the classification of spectral shapes, radiation mechanism models, and the influence of the ISM. The diversity of spectral shapes observed in pulsars reflects the intricate nature of their radiation mechanisms. This paper identifies several key spectral phenomena, such as low-frequency and high-frequency turnovers, as well as high-frequency cutoffs, and investigates the underlying physical processes responsible for these features. The low-frequency turnover is found to be influenced by both the radiation mechanisms and the effects of the ISM, consistent with the predictions of the coherent curvature radiation model. 

In terms of radiation mechanisms, synchrotron and curvature radiation are the predominant models across various frequency ranges; however, a unified model that bridges the two remains absent. The correlation between spectral indices and pulsar characteristics underscores the substantial influence of rotational energy loss rates and magnetospheric structure on radiation efficiency. Additionally, the interactions between different radiation regions and particle acceleration processes warrant further exploration to enhance our understanding of pulsar emission physics.

The effects of the ISM significantly influence the propagation of pulsar signals, including dispersion, scattering, and FFA, especially pronounced in the low-frequency regime. Therefore, understanding these effects is crucial for the accurate measurement of pulsar spectral features. Recent advances in low-frequency radio observational technologies, such as LOFAR and MWA, have expanded the pulsar spectral dataset and facilitated in-depth studies of radiation mechanisms and the distribution of interstellar plasma. These technological breakthroughs provide a solid foundation for future observations using next-generation radio telescopes, such as the SKA.

Looking ahead, as radio telescope technologies advance and observational methods improve, pulsar spectral research will encounter both opportunities and challenges. Low-frequency observations continue to face technical constraints such as ionospheric interference and Galactic background noise; however, these observations gradually reveal insights into pulsar radiation mechanisms and the ISM characteristics.

\section*{Acknowledgments}
This work is partially supported by the National SKA Program of China (No. 2020SKA0120201) and the National Key Research and Development Program of China (No. 2022YFC2205203), the Major Science and Technology Program of Xinjiang Uygur Autonomous Region (No. 2022A03013-1), the National Natural Science Foundation of China (NSFC grant Nos. 12303053). Z.G.W. is supported by the Youth Innovation Promotion Association of CAS under No. 2023069, and the Tianshan Talent Training Program (No. 2023TSYCCX0100).  




\bibliography{Wiley-ASNA}%

\end{document}